# Investigating phase transitions from local crystallographic analysis based on machine learning of atomic environments

Rama K. Vasudevan,[1] Maxim Ziatdinov,[1] Lukas Vlcek,[3,4] Anna N. Morozovska,[2] Eugene A. Eliseev,[5] Shi-Ze Yang,[6] Yongji Gong,[7] Pulickel Ajayan,[8] Wu Zhou,[1,a] Matthew F. Chisholm,[1] and Sergei V. Kalinin[1,b]

[1] The Center for Nanophase Materials Sciences, Oak Ridge National Laboratory, Oak Ridge, TN 37831
[2] Institute of Physics, National Academy of Sciences of Ukraine, 46, pr. Nauky, 03028 Kyiv, Ukraine
[3] Materials Science and Technology Division, Oak Ridge National Laboratory, Oak Ridge TN 37831, USA
[4] Joint Institute for Computational Sciences, University of Tennessee, Knoxville, Oak Ridge, TN 37831, USA
[5] Institute for Problems of Materials Science, National Academy of Sciences of Ukraine, Krjijanovskogo 3, 03142 Kyiv, Ukraine
[6] Center for Functional Nanomaterials, Brookhaven National Laboratory, NY
[7] School of Materials Science and Engineering, Beihang University, Beijing 100191, China
[8] Department of Materials Science and Nano Engineering, Rice University, Houston, Texas 77005, USA

**Abstract**

Traditionally, phase transitions are explored using a combination of macroscopic functional characterization and scattering techniques, providing insight into average properties and symmetries of the lattice but local atomic level mechanisms during phase transitions generally remain unknown. Here we explore the mechanisms of a phase transition between the trigonal prismatic and distorted octahedral phases of layered chalogenides in the $MoS_2$ – $ReS_2$ system from the observations of local degrees of freedom, namely atomic positions by Scanning Transmission Electron Microscopy (STEM). We employ local crystallographic analysis based on machine learning of atomic environments to build a picture of the transition from the atomic level up and determine local and global variables controlling the local symmetry breaking. In particular, we argue that the dependence of the average symmetry breaking distortion amplitude on global and local concentration can be used to separate local chemical and global electronic effects on transition. This approach allows exploring atomic mechanisms beyond the traditional macroscopic descriptions, utilizing the imaging of compositional fluctuations in solids to explore phase transitions over a range of realized and observed local stoichiometries and atomic configurations.

[a] Present address: School of Physical Sciences and CAS Center for Excellence in Topological Quantum Computation, University of Chinese Academy of Sciences, Beijing 100049, China
[b] Corresponding author, sergei2@ornl.gov

Phase transitions are a central topic in condensed matter physics, given their prime importance not only for specific material applications, but also in phenomena such as superconductivity,[1, 2] relaxor ferroelectricity,[3, 4] and quantum spin-liquid[5, 6] behaviors. Depending on the chemical composition, temperature, pressure, or electrostatic field effect doping, materials evolve complex phase diagrams. The enhanced susceptibilities and giant responses in the vicinity of phase transitions attract the attention of applied scientists, with examples ranging from giant dielectric susceptibilities[7, 8] and piezoelectricity in ferroelectrics to phase-change field effect transistors based on metal-insulator and transitions or ferroelectric switching,[9] and many others.[10]

The mechanisms of phase transitions have been explored at multiple levels of theoretical description. At the most fundamental level, density-functional theory provides insight into electronic and structural instabilities that can drive the transition between dissimilar phases, that can be further augmented via effective Hamiltonian methods[11-13] to provide information on finite-temperature properties. A panoply of lattice models enabled the exploration of collective phenomena in systems of large numbers of interacting spins on a predefined lattice.[14-16] Finally, group theory and closely linked to it phase-field theory describes phase transitions on the mesoscopic level via the dynamics of effective order parameter fields, the nature of which is intrinsically linked to the symmetry of the related phases.[17]

The traditional experimental counterparts to these are the combination of macroscopic property measurements and scattering techniques. The former provide insight into the macroscopically averaged functionalities determined by the property of the phases, where the character of anomalies (e.g. critical exponents and dynamics) in the transition region are intrinsically linked to the nature of the transition. Scattering techniques provide the information on the basic symmetry of the phases, and under certain conditions offer insight into the nature of interactions driving the transition. However, the broad set of phenomena that evolve on the atomic and mesoscopic scales, ranging from exact atomistic mechanisms to collective phenomena such as formation of the spatially non-uniform states and critical fluctuations to formation and dynamics of topological defects remain poorly understood and accessible only on average, via long-wavelength scattering. This limitation is particularly stringent for systems with symmetry incompatible or frustrated interactions that often lead to spatially inhomogeneous ground states, poorly amenable to classical scattering studies.[18, 19]

For many materials classes, breakthroughs in understanding the mesoscopic nature of phase transitions were achieved with the introduction of local mesoscopic probes, including piezoresponse force microscopy[20] for ferroelectric materials, magnetic force microscopy[21] for ferromagnets and superconductors, and focused X-ray[22] for structural transitions. These methods allowed visualization of topological defects, intrinsically linked to the nature and symmetries of the order parameter. At the atomic level, STEM has provided insight into the nature of ferroic states and ferroic behaviors at topological defects in a broad set of ferroelectric, ferroelastic, and chemically ordered systems via observations of atomic scale displacements and strains.[23-25] However, the vast majority of these studies are limited by the observations at fixed temperature and composition, in which case relevant information can be determined only from the behaviors at topological and structural defects.[26] Similarly, virtually all analyses to date utilized the macroscopic descriptors derived from structural information such as polarization or octahedral tilts extracted from the atomic positions, ignoring the multiple degrees of freedom observable from atomically resolved image.

Recently, we proposed that structures in solids can be defined from the bottom up via the statistical analysis of the local atomic neighborhoods and structures.[27] Here, we apply this approach to explore a phase transition between the trigonal prismatic and distorted octahedral phases in the $MoS_2$ – $ReS_2$ solid solution system, establishing the relationship between local structural degrees of freedom and chemical structure. We demonstrate that the phase transition can be determined as a result of the symmetry breaking at a single atom level via analysis of relevant statistical modes identified by the principal component analysis (PCA). The compositional fluctuations and more generally multiple atomic configurations present on the atomic level hence allow to explore local mechanisms for doping-induced symmetry breaking on the single-atom level and extrapolated over a large composition space. We further argue that exploring the magnitude of symmetry breaking distortions as a function of both local and global composition may potentially allow decoupling of the short-range chemical and long-range electronic effects in materials.

**$MoS_2$-$ReS_2$ Imaging**

The $MoS_2$-$ReS_2$ system offers convenient model system for exploring structural transitions.[28, 29] Here, $MoS_2$ exists in the trigonal prismatic semiconducting H phase under normal conditions. The point group of unstrained bulk $MoS_2$ unit cell is $\bar{6}m2$ (trigonal prismatic).[30] This aristotype structure allows for multiple displacive structural transformations, permitting rapid and possibly reversible switching between semiconducting H, and metallic or semi-metallic (T') crystal structures.[31] The $\bar{6}m2$ point group is reduced to $6mm$ point group for the single-layer $MoS_2$ (SL-$MoS_2$), because all "out-of-plane" symmetry operations moving atoms away from the 2D-plane disappear.[32] The H phase can be transformed to a metallic octahedral (T) phase by Re doping,[33] though the T phase is metastable without the doping or other specific treatment. Under the specific treatment H phase can also be transformed to the distorted octahedral phase with clusterization of metal atoms into zigzag chains (zigzag-shape phase "Z").[34] On the other hand, $ReS_2$, exist in the diamond-shape (DS) phase with triclinic symmetry, where the neighboring Re clusters are linked along the b[010] axis to form Re DS-chains.[35] The point group of unstrained bulk $ReS_2$ unit cell is $3m$ (triclinic).[36, 37] The $3m$ point group remains unchanged (as containing no "out-of-plane" symmetry operations) and stable under normal conditions for the SL-$ReS_2$. Lin et al.[38] showed here that SL-$ReS_2$ is a stable *n*-type semiconductor which exhibits an order of magnitude difference in anisotropic conduction, unlike most 2D materials. Note that most $MoS_2$ and $ReS_2$ structures cannot coexist without defects and broken bonds,[39] offering an ideal model case for symmetry-incompatible phases.

To obtain insight into the structural properties of these materials, the structure was explored via STEM as described in Ref [[40]]. These studies provided the insight into the structure and cation distribution in this system and allowed us to establish the thermodynamics defining the cation distribution in the system, in particular the weakly repulsive interactions between similar cations favoring the formation of an almost ideal solid solution. Here, we study the evolution of the structure across the solid solution series, in particular the evolution of the cationic neighborhoods on transition from $MoS_2$ to $ReS_2$. We derive the nature of the distortion mode that drives phase transition, and further use this information to analyze the role of local concentrations (short-range chemistry) and global factors (average concentration) that drives phase transition.

The atomic resolution images across the composition series are shown in Fig. 1, with the corresponding 2D Fast Fourier Transforms (FFT) shown below. The $MoS_2$ phase shows well-formed hexagonal lattice of pure H phase, as expected. On alloying with Re, the system starts to develop characteristic modulations. This process is associated with the broadening and splitting of the peaks on the FFT and disorder in the system. Finally, the purely $ReS_2$ phase develops the characteristic ordering pattern expected for the triclinic phase of 2D chalcogenides. The disorder in this case is decreased as can be observed from the 2D FFT and neighborhood histogram images, where the peak splitting indicates the symmetry lowering compared to the H phase and peak width being indicative of the degree of disorder. Note that this information is derived as a global descriptor from the atomically resolved STEM image and is roughly equivalent to what could be obtained from scattering measurements. As such, it does not contain element-specific or local information.

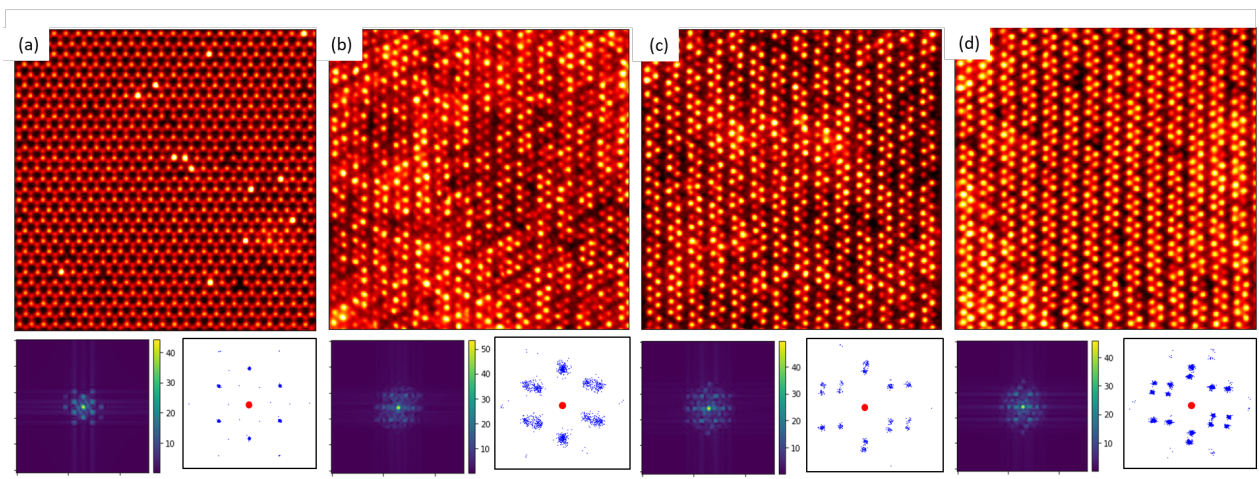

**Figure 1.** High-resolution scanning transmission electron microscopy images of $Mo_{1-x}Ru_xS_2$ with x = 0.05, 0.55, 0.78, and 0.95 (a-d). Also shown as the corresponding Fast Fourier Transforms and the atomic neighborhood histograms. The data used is the same as in Ref. [40].

**Local Crystallography Analysis**

To obtain insight into the local phase transition mechanisms, we analyze the structure of each composition at the atomic level. We utilize a workflow based on rough determination of atomic position (indexing) with subsequent refinement via 2D Gaussian fitting to each atom to establish the position of the cations in the material with sub-pixel accuracy.[41] The indexing was performed using the combination of AICrystallographer deep learning tools and a motif-matching algorithm available via the open source PyCroscopy package.[42-44] The identity of the cations were established based on local contrast, with the threshold intensity between Mo and Re adjusted based on expected bulk composition. Note that while S atoms and correspondingly S vacancies can be unambiguously located in the $MoS_2$, this is not the case for other compositions. Given the relatively small number of sulfur vacancies, the effects are ignored in the subsequent analysis. The original data, the codes used for data analytics and the full analysis workflow, as well as multiple additional representations of the structures are available as a Jupyter notebook that allow the interested reader to reproduce the analysis in this paper in full.

To gain insight into the structure of the solid solutions, we utilize the local crystallography approach.[45] Here, for each cation, we define the nearest atomic neighborhood of $N$ atoms. Given the symmetry properties of the host lattice and the absence of defects that break chemical bonding pattern, here we explored $N = 6, 12, 24$, even though other numbers can be explored. Each atom is then described by the $2N$ component vector defining its nearest neighbor structure (i.e., $(x, y)$ positions of the neighbors relative to the selected atom). Note that the descriptor can be adapted for a specific problem, i.e. can be decomposed into purely distance information (closely linked to chemical bond disorder), angular part (an approach adopted in simple liquid theory when defining hexatic and tetratic order parameters), adapted to allow for rotational invariance, or for specific lattice symmetries. Here, we use the most general description where the full $2N$ vector is used to describe the local crystallographic structure. It is important to note that this analysis is performed separately for Mo and Re atoms, i.e. is chemically-specific from the beginning. Furthermore, the atomic identities of neighborhood atoms are also preserved and can be used to describe local chemical composition, as will be described below.

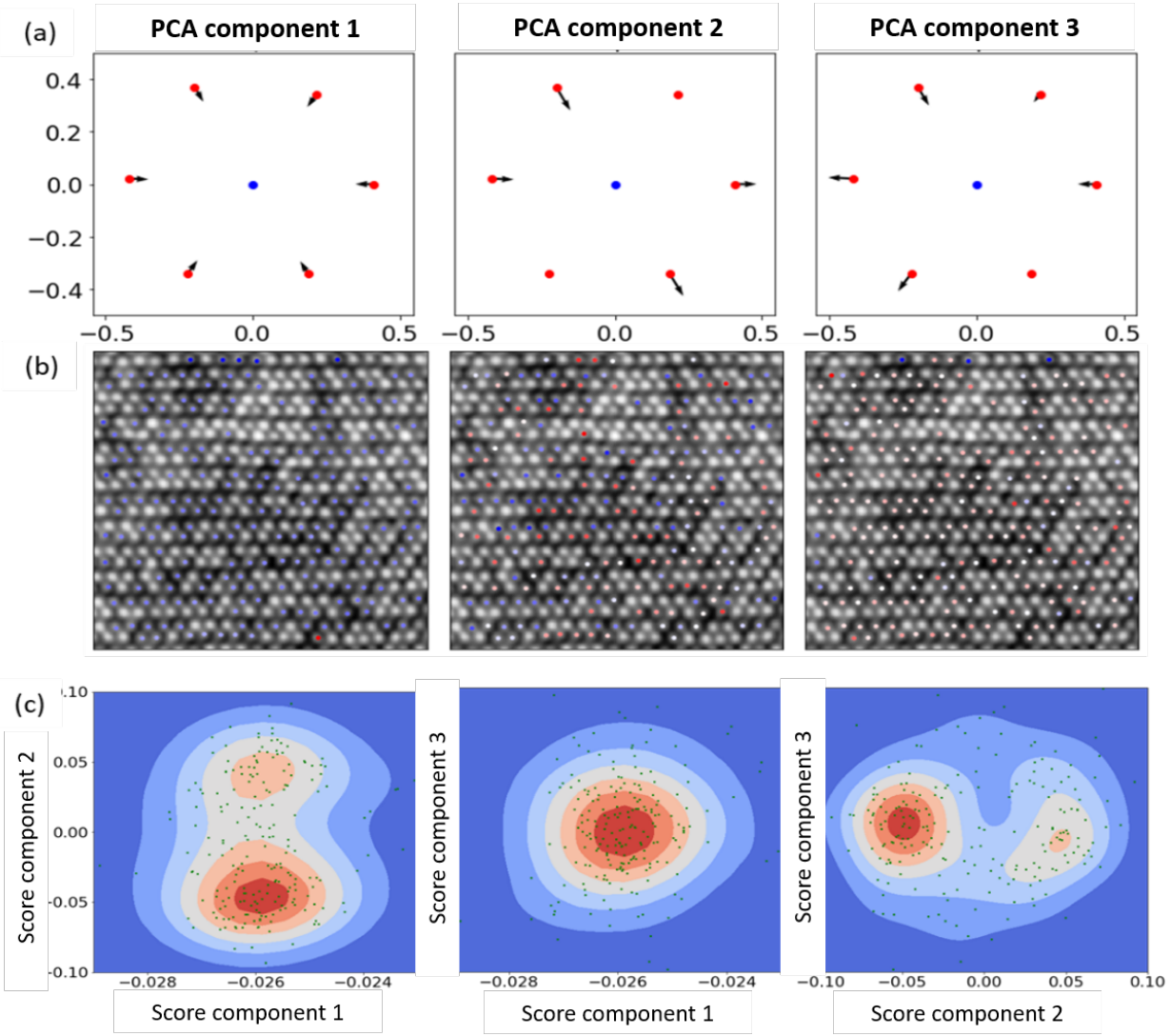

**Figure 2.** Principal component analysis of the $Mo_{0.45}Re_{0.55}S_2$ image. (a) Atom displacements with respect to the central atom corresponding to the first three principal components, (b) the associated maps of PCA scores with red being positive and blue being negative, specifically for Mo atom neighborhoods. (c) Joint density distributions of the component scores for the first three components.

The results of the crystallographic analysis of the $Mo_{0.45}Re_{0.55}S_2$ phase are shown in Fig. 2. Here, the atomic neighborhood vector is analyzed using the PCA, separating statistically significant orthogonal distortions for the lattice. The PCA components shown in Fig. 2 provide insight into the local fluctuations of atomic structure in the material from a statistical perspective. Note that while PCA is a purely an information theory-based tool, the behavior of the PCA components can nonetheless readily offer physical interpretation. Here, the first PCA component represents the isotropic compression or expansion around the central atom (depending on the sign of the PCA loading), and represents the classical Vegard strain behavior associated with the difference with molar volumes between Mo and Re. The second PCA component represents a deformation of atomic neighborhood where two atoms move closer to the central atom and two are moving further away, while maintaining symmetry with respect to the crystallographic axis. This mode represents the dominant symmetry breaking mode for all studied compositions and may provide a locally defined order parameter, as will be describe further. The third component represents more complex deformations of the unit cell and given the limited statistics and uncertain concentration of sulfur vacancies are not analyzed for individual compositions. The complementary analysis via k-means clustering is provided in Supplementary Analysis.

Note the two important aspects of this local crystallographic analysis. First, the multivariate analysis of local crystallography separates the multitude of structural distortions for specific local atomic configurations into the statistically relevant components, providing a full description of statistical behaviors associated with local symmetry breaking and the phase transition. This analysis is based only on the information contained in the image, and therefore is not biased by *a priori* assumptions of any physical model. Secondly, the analysis can be performed element specific, i.e. the structure of the chemical neighborhoods and their evolution across the phase transition can be performed for Mo and Re separately. While the large number of possible components can result in a significant human bias in selection of relevant representations, we provide a Jupyter notebook that allows the readers to explore the data representations at will and contains the results for all the compositions (and can be used for analysis of the reader's data as well).

The individual images in Figure 1 contain both the chemical (local atomic identities) and physical (distortions from H structure) information across the composition series. We note however that while from the global perspective we explore only 4 compositions, the situation is different from the local perspective, since each image contains multiple local atomic configurations. For example, if we assume that the material is the ideal solid solution with overall Re concentration $x$, the probability of a local composition with $i$ Re atoms within a cluster of $n = 7$ atoms will be defined by the binomial distribution $P(i, n, x) = \binom{n}{i} x^i (1-x)^{n-i}$. From this perspective, each sample will have regions with different compositions $Mo_{1-i/7}Re_{i/7}$, $i = 0, .., 7$ with number of these regions determined by overall concentration. For non-ideal solid solutions, the probabilities are dependent on the segregation energy and in principle the latter can be determined from observed atomic configurations. Here we however utilize the fact that individual atomic configurations are already present in the images, and hence we can explore the physical and chemical information contained at each lattice site. In this manner, the atomically resolved images can be identified with the combinatorial library of composition-structural distortion relationships. The data across multiple compositions then allows to improve the statistics for different neighborhoods and allows for exploration of self-consistency. Note that this logic follows that of the macroscopic Ginzburg-Landau theory, where the order parameter expansion is expected to be constant across the phase transition, and only certain term(s) changes sign.

**Distortion Modes with Local and Global Chemistry**

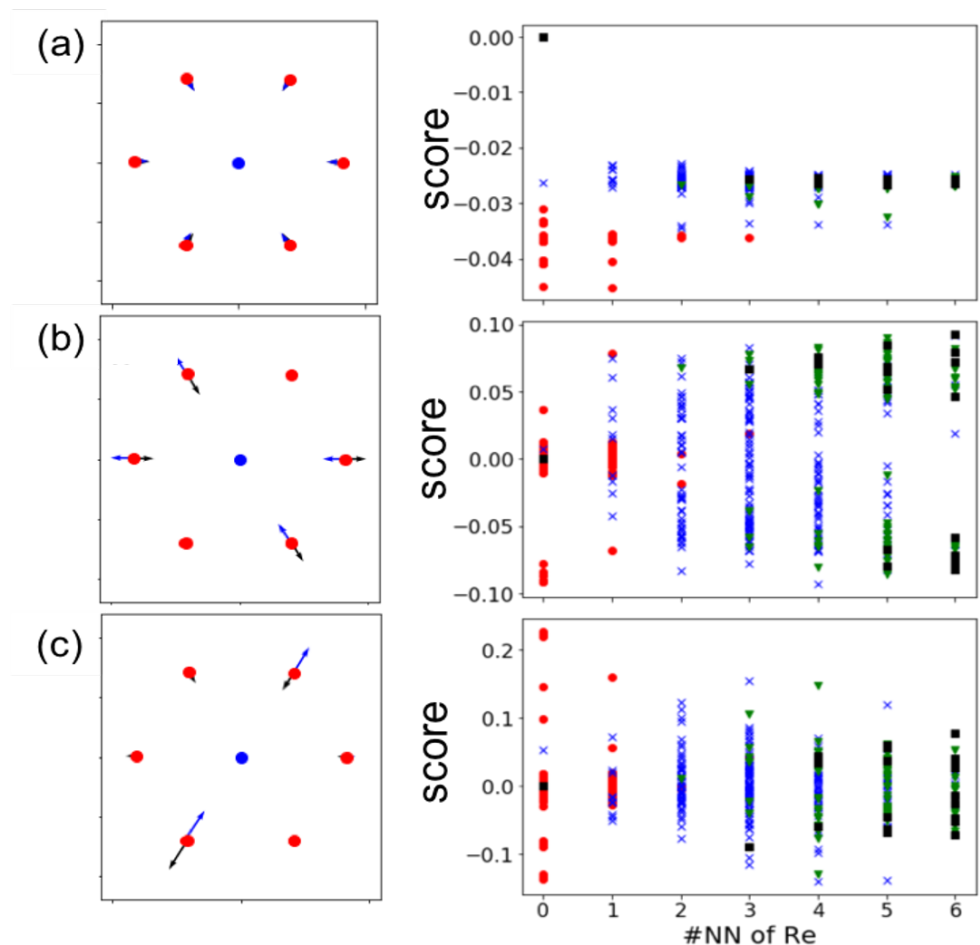

**Figure 3. Evolution of statistical principal components across the composition series.** (a-c) Principal components (left) and the corresponding scores (right) for the (a) 1st, (b) 2nd and (c) 3rd principal component for Mo and Re neighbors. On the left in each panel are shown first three components determined across the full phase diagram (black and blue arrows for Mo and Re neighborhoods, respectively). On the right, distributions of the corresponding component scores as a function of local chemical composition (i.e. number of neighbors of Re), individually colored for each composition that the statistic (red, blue, green and black for $x=0.05$, $x=0.55$, $x=0.78$ and $x = 0.95$, respectively). Note that while the 1st component is approximately uniform, the 2nd component shows clear splitting for 2 and larger number of neighbors of Re.

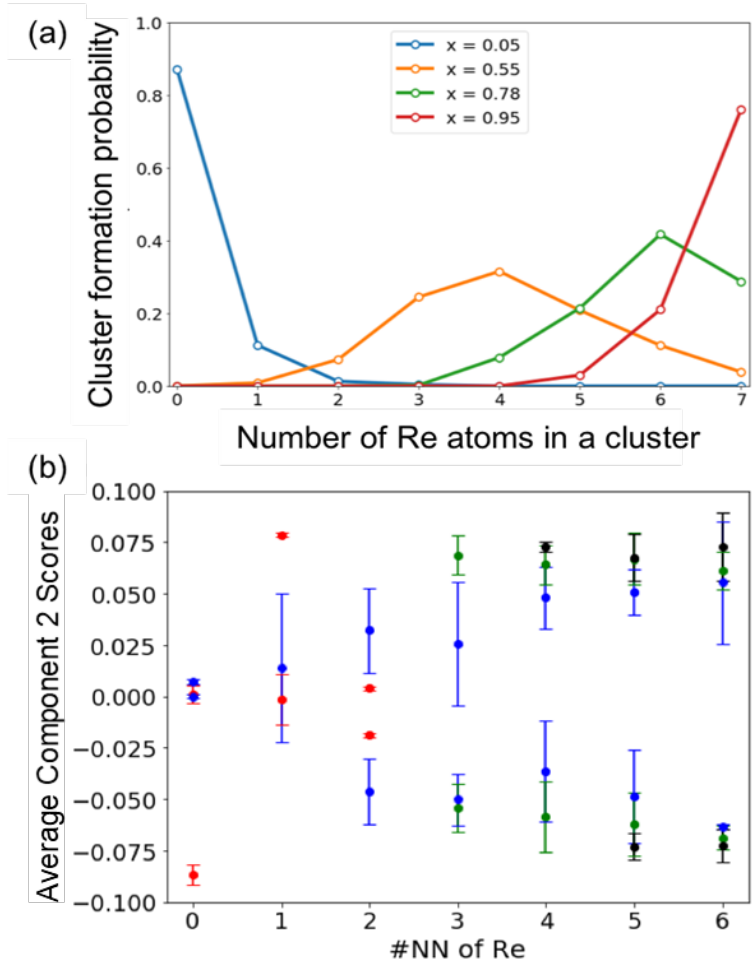

**Figure 4. Local analysis of Re neighbors.** (a) The probability of clusters comprised of a central atom and 6 nearest neighbors as a function of Re fraction $x$. (b) Gaussian mixture approximation for the distribution of the second PCA component scores as a function of the number of Re nearest neighbors of a central Re atom. The error bars indicate the standard deviation of the two-component Gaussian mixture model; the colors refer to different global Re fractions, $x$, using the same color scheme as in Fig. 3. Comparison of the two plots illustrates that the splitting, representing local structural transition, depends on the local, but not global, composition.

Next, we aim to explore the relationship between the chemical composition fluctuations and the phase transition mechanisms. Here we note that $MoS_2$-$ReS_2$ system offers a very complex case for such a study, since the phase transition occurs at the ~50% alloying and the solid solution is close to ideal. Correspondingly, characteristic separations between dissimilar atoms is comparable to the lattice parameter, precluding clear separation of structural distortions around individual atoms as would be the case for very diluted solid solution. More generally, both local chemical effects and global electronic effects can give rise to the structural evolution. Here, we aim to distinguish the two.

As a measure of local chemical composition, we choose a number of Mo/Re neighbors for a chosen central atom (also Mo or Re). As a measure of symmetry-breaking, we choose the values of the PCA scores (*i.e.*, values projected onto the principal component) associated with a given atom, where the statistical analysis is performed over the full composition series. Note that while more complex descriptors based on relative atomic positions, sub-clusters, etc. can be chosen, the requirements for statistics progressively grow for more complex structural elements, and hence we defer these studies for the future. Similarly, this analysis can be generalized by exploring the intensity of Fourier components associated with specific statistical modes of lattice distortions to explore the measure of long-range order; similarly, we defer this to the future studies and here concentrate only on local descriptors.

The relationship between local chemical composition and symmetry-breaking behavior is shown in Figures 3 and 4. Here, the distributions of the PCA components for Re central atoms are shown as a function of the number of Re neighbors with individual points colored by the composition they arise from. The first component shows relatively uniform distribution with separate behavior at the $x = 0.05$ composition. The third component shows relatively broad behavior as a function of the number of Re atoms in the neighborhood, with the maximal width for 2 and 3 neighbors. Finally, the most interesting behavior is observed for the second PCA component. Here, for 0 and 1 neighbors of Re, the distribution of the principal component scores is unimodal. It broadens for the 2 neighbors and becomes bimodal for the 4,5 and 6 neighbors. These behaviors hold independently of the global compositions as can be seen by the blue, green and black points overlapping. We believe that this behavior represents the local phase transition mechanism resolved at the atomic level. In other words, the same distortions are seen regardless of the composition but depend only on local environment, specifically the number of neighbors of Re.

We can in fact make a machine learning model to predict the 2$^{nd}$ PCA component scores. For this purpose, we utilized a Gaussian mixture model (GMM)[46], which is a probabilistic model that assumes datapoints are generated from some finite number of Gaussian distributions, in this case 2 (this was found using the Bayesian information criterion). We fit a GMM to each set of data (*C*,*N*) where C is the composition and N is the number of nearest neighbors of Re. Thus in total we have C*N total GMMs. The result of the GMM analysis is shown in Fig. 4(b) where the means and variances are plotted. This model appears to capture the trends shown in Fig. 3(b), and the overlap of the predicted means for the different colors indicates that local chemical character, not global composition, dictates the presence of this distortion.

In general, we note that the phase transition in the material can be assumed to be mediated by short- and long-range mechanisms. The short-range mechanisms are related to the bonding and molar volume effects, that under certain conditions can give rise to the global transition. As such, these are sensitive to ordering and positions of local atomic units. The long-range effects, e.g. those related to the changes in the average electronic concentration and structural instability due to e.g. Fermi surface nesting are determined by average dopant concentrations. Note that while intrinsically mixed in any real material, these can be separated via external interventions. For example, pressure effects (via differential of compressibility) will affect the local chemical effects, whereas electronic doping via e.g. gating will affect the electronic driven instabilities but not chemical effects.

To establish the role of the local (neighborhood of a single atom) and global composition on the phase transition, the principal component score distributions (specifically Fig. 3(b)) were further analyzed using a Gaussian Process (GP) regression model.[47-50] That is, we can use a more advanced machine learning model than a simple GMM, and hopefully derive more insight into the mechanism. Here, GP, a Bayesian machine learning model, is used as a universal interpolator, reconstructing function $f$ over set of arguments and values $D = \{(x_1, y_1), \ldots (x_N, y_N)\}$. Here, it assumed that each pair is related by $y = f(x) + \varepsilon$, where $\varepsilon$ is white noise, and the function $f$ has a prior distribution $f \sim \mathcal{GP}(0, K_f(x, x'))$, where $K_f$ is a covariance function (kernel).[47] The covariance matrix of the GP posterior distribution serves as an estimator of the uncertainty in the interpolation. We explored the GP interpolation both as function of local concentration only, with global concentration being the free parameter, and as a function of both local and global concentration.

The interpolated values of the symmetry breaking mode (i.e., the 2nd PCA component scores) for local concentration only are shown in Suppl. Mat. (Fig. S3). In that figure, the magnitude of the symmetry-breaking mode increases with local concentration. For the 5% composition only a small number of compositional variants are available. At the same time, for the three other compositions the distortion increases with local composition and saturates at the same value, suggesting the universality of the symmetry breaking phenomena as a function of local composition.

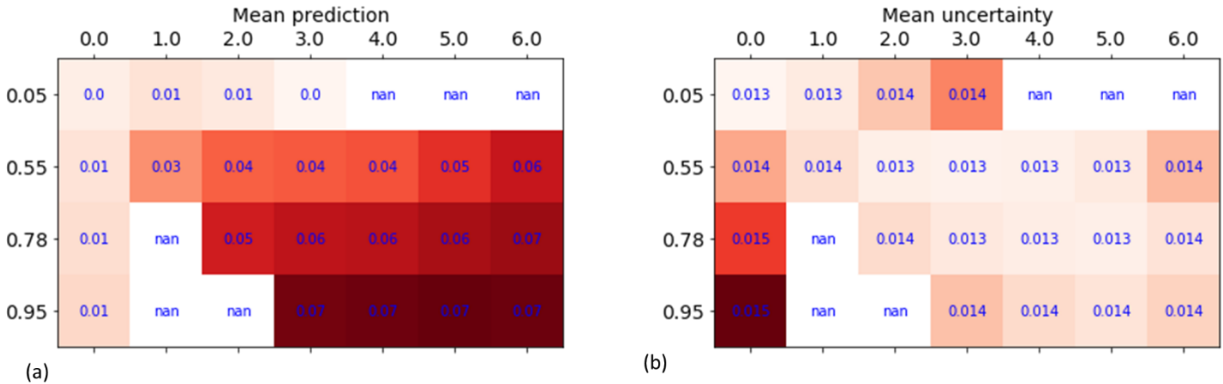

**Figure 5. Global and local effects on symmetry breaking.** (a) Gaussian Processing averaged prediction of second PCA component score as a function of global (vertical axis) and local (horizontal) composition. (b) Associated mean uncertainty.

To obtain further insight regarding this behavior and potentially decouple the local chemical composition effects and global concentration effects, we perform the GP regression on local distortion as a function of local and global composition, $D(N, x)$. The resulting prediction matrix and associated deviations are shown in Figure 5. Note that while this matrix allows for multiple possible interpretations, here we focus on the separation of short-range chemical effects controlled by local neighborhood and long-range effects. The direct examination of the Figure 5 illustrates that the dependence of the symmetry breaking on the local composition shows more pronounced effect then global composition. An estimate via linear regression $D(N, c) = aN + b\,x + d$ yields the ration of short to long range effects as $b/(6\,a) = 0.5$, suggesting the preponderant role of local chemical effects on phase transition.

To summarize, we have explored the nature of the phase transition between the trigonal prismatic and distorted octahedra phases of layered chalcogenides exemplified by the alloying in the $MoS_2$ – $ReS_2$ system. Based on the element specific local crystallographic analysis, we use PCA to explore the evolution of the chemical neighborhood and structural symmetry breaking via statistical mode analysis and correlation between local physical descriptors and chemical composition. We observe that the magnitude of the second principal component can be used as a statistically significant descriptor that describes symmetry breaking and phase transition. We note that this behavior is local and is determined and defined on the level of individual atomic neighborhoods visualized via high-resolution STEM.

We believe that this approach provides a bottom-up pathway to explore the nature of concentration induced phase transitions in condensed matter systems. While the further developments of this approach will require increased statistics to collect the information relevant to relevant defects, it offers a universal approach to study phase transitions in low dimensional systems. This approach is also applicable for 3D materials if the structural distortions are isotropic in the beam direction, as is often the case for ferroelastic and ferroelastic materials. The analysis can be considerably simplified if the transition happens at lower concentrations of dopants allowing for increasing the separability between chemical and physical characteristic length.

The method proposed here can further be extended for analysis of the thermodynamics of the phase transition from the observed distortions, as has been previously demonstrated for atomic structure. Similarly, new opportunities emerge in targeted probing of materials functionalities by combining these analysis tools with the electron beam-based approaches to engineer cation sublattice via atomic motion[28, 51-53] and vacancy formation.

**Acknowledgements**

The effort at ORNL including electron microscopy (S.Z.Y., M.F.C., W.Z.) and image analytics (R.V.K., S.V.K., L.V.) was supported by the U.S. Department of Energy, Office of Science, Basic Energy Sciences, Materials Sciences and Engineering Division and performed at the ORNL Center for Nanophase Materials Sciences, which is a US DOE Office of the Science User Facility.

**Conflict of Interest Notice**

The authors declare no conflict of interest.

**Data Availability**

All data and analysis are included with this manuscript via the Jupyter notebook.

**Methods**

*Sample growth:*

Molybdenum oxide powder (99%, Sigma Aldrich), sulfur powder (99.5%, Sigma Aldrich) and ammonium perrhenate (99%, Sigma Aldrich) were used as precursors for CVD growth. A selected ratio of molybdenum oxide and ammonium perrhenate was added to an alumina boat with a Si/SiO2 (285 nm) wafer cover. The furnace temperature was ramped to 550 °C in 15 min and then kept at 550 °C for another 15 min for the growth of the RexMo1-xS2 alloy materials. Sulfur powder in another alumina boat was placed upstream where the temperature was roughly 200 °C. After growth, the furnace was cooled to room temperature using natural convection. The growth process was carried out with 50 SCCM argon at atmospheric pressure.

*Electron microscopy characterization:*

The $Re_xMo_{1-x}S_2$ flakes were transferred to TEM grids by spin coating PMMA to support the flakes and etching with KOH to release them from the substrates (by dissolving the SiO2). The annular dark-field images (ADF) were collected using a Nion UltraSTEM100 microscope operated at 60 kV. The as-recorded images were filtered using a Gaussian function (full width half maximum = 0.12 nm) to remove high-frequency noise. The convergence half angle of the electron beam was set to 30 mrad and the collection inner half angle of the ADF detector was 51 mrad. The samples were baked in vacuum at 140 0C overnight before STEM observation. During STEM observation, the probe current was controlled between 10 pA to 60 pA to reduce beam damage.

**Analysis**

A full Jupyter notebook outlining the analysis is included with the manuscript and can be run online on Google Colab (colab.research.google.com).

# References

1. Keimer, B.; Moore, J. E., The physics of quantum materials. *Nat. Phys.* **2017,** *13* (11), 1045-1055.
2. Tokura, Y.; Kawasaki, M.; Nagaosa, N., Emergent functions of quantum materials. *Nat. Phys.* **2017,** *13* (11), 1056-1068.
3. Westphal, V.; Kleemann, W.; Glinchuk, M. D., DIFFUSE PHASE-TRANSITIONS AND RANDOM-FIELD-INDUCED DOMAIN STATES OF THE RELAXOR FERROELECTRIC PBMG1/3NB2/3O3. *Physical Review Letters* **1992,** *68* (6), 847-850.
4. Glazounov, A. E.; Tagantsev, A. K.; Bell, A. J., Evidence for domain-type dynamics in the ergodic phase of the PbMg1/3Nb2/3O3 relaxor ferroelectric. *Physical Review B* **1996,** *53* (17), 11281-11284.
5. Tokura, Y.; Nagaosa, N., Orbital physics in transition-metal oxides. *Science* **2000,** *288* (5465), 462-468.
6. Imada, M.; Fujimori, A.; Tokura, Y., Metal-insulator transitions. *Rev. Mod. Phys.* **1998,** *70* (4), 1039-1263.
7. Damjanovic, D., Ferroelectric, dielectric and piezoelectric properties of ferroelectric thin films and ceramics. *Reports on Progress in Physics* **1998,** *61* (9), 1267-1324.
8. Zeches, R. J.; Rossell, M. D.; Zhang, J. X.; Hatt, A. J.; He, Q.; Yang, C. H.; Kumar, A.; Wang, C. H.; Melville, A.; Adamo, C.; Sheng, G.; Chu, Y. H.; Ihlefeld, J. F.; Erni, R.; Ederer, C.; Gopalan, V.; Chen, L. Q.; Schlom, D. G.; Spaldin, N. A.; Martin, L. W.; Ramesh, R., A Strain-Driven Morphotropic Phase Boundary in BiFeO(3). *Science* **2009,** *326* (5955), 977-980.
9. Miller, S. L.; McWhorter, P. J., PHYSICS OF THE FERROELECTRIC NONVOLATILE MEMORY FIELD-EFFECT TRANSISTOR. *Journal of Applied Physics* **1992,** *72* (12), 5999-6010.
10. Waser, R., Nanoelectronics and Information Technology *Nanoelectronics and Information Technology* **2012**.
11. Kingsmith, R. D.; Vanderbilt, D., THEORY OF POLARIZATION OF CRYSTALLINE SOLIDS. *Physical Review B* **1993,** *47* (3), 1651-1654.
12. Lai, B. K.; Ponomareva, I.; Naumov, II; Kornev, I.; Fu, H. X.; Bellaiche, L.; Salamo, G. J., Electric-field-induced domain evolution in ferroelectric ultrathin films. *Physical Review Letters* **2006,** *96* (13).
13. Ponomareva, I.; Naumov, II; Kornev, I.; Fu, H.; Bellaiche, L., Atomistic treatment of depolarizing energy and field in ferroelectric nanostructures. *Physical Review B* **2005,** *72* (14).
14. Kamiya, Y.; Kato, Y.; Nasu, J.; Motome, Y., Magnetic three states of matter: A quantum Monte Carlo study of spin liquids. *Phys. Rev. B* **2015,** *92* (10).
15. Wang, Y.-C.; Zhang, X.-F.; Pollmann, F.; Cheng, M.; Meng, Z. Y., Quantum Spin Liquid with Even Ising Gauge Field Structure on Kagome Lattice. *Phys. Rev. Lett.* **2018,** *121* (5), 057202.
16. Becker, J.; Wessel, S., Diagnosing Fractionalization from the Spin Dynamics of $ {Z}_{2}$ Spin Liquids on the Kagome Lattice by Quantum Monte Carlo Simulations. *Phys. Rev. Lett.* **2018,** *121* (7), 077202.
17. Steinbach, I., Phase-Field Model for Microstructure Evolution at the Mesoscopic Scale. *Annual Review of Materials Research* **2013,** *43* (1), 89-107.
18. Keen, D. A.; Goodwin, A. L., The crystallography of correlated disorder. *Nature* **2015,** *521* (7552), 303-309.
19. Senn, M. S.; Keen, D. A.; Lucas, T. C. A.; Hriljac, J. A.; Goodwin, A. L., Emergence of Long-Range Order in BaTiO3 from Local Symmetry-Breaking Distortions. *Physical Review Letters* **2016,** *116* (20).
20. Gruverman, A.; Auciello, O.; Ramesh, R.; Tokumoto, H., Scanning force microscopy of domain structure in ferroelectric thin films: imaging and control. *Nanotechnology* **1997,** *8*, A38-A43.
21. Martin, Y.; Wickramasinghe, H. K., MAGNETIC IMAGING BY FORCE MICROSCOPY WITH 1000-A RESOLUTION. *Appl. Phys. Lett.* **1987,** *50* (20), 1455-1457.

22. Winarski, R. P.; Holt, M. V.; Rose, V.; Fuesz, P.; Carbaugh, D.; Benson, C.; Shu, D. M.; Kline, D.; Stephenson, G. B.; McNulty, I.; Maser, J., A hard X-ray nanoprobe beamline for nanoscale microscopy. *J. Synchrot. Radiat.* **2012,** *19*, 1056-1060.

23. Jia, C. L.; Nagarajan, V.; He, J. Q.; Houben, L.; Zhao, T.; Ramesh, R.; Urban, K.; Waser, R., Unit-cell scale mapping of ferroelectricity and tetragonality in epitaxial ultrathin ferroelectric films. *Nat. Mater.* **2007,** *6* (1), 64-69.

24. Borisevich, A. Y.; Chang, H. J.; Huijben, M.; Oxley, M. P.; Okamoto, S.; Niranjan, M. K.; Burton, J. D.; Tsymbal, E. Y.; Chu, Y. H.; Yu, P.; Ramesh, R.; Kalinin, S. V.; Pennycook, S. J., Suppression of Octahedral Tilts and Associated Changes in Electronic Properties at Epitaxial Oxide Heterostructure Interfaces. *Physical Review Letters* **2010,** *105* (8).

25. Jang, J. H.; Kim, Y. M.; He, Q.; Mishra, R.; Qiao, L.; Biegalski, M. D.; Lupini, A. R.; Pantelides, S. T.; Pennycook, S. J.; Kalinin, S. V.; Borisevich, A. Y., In Situ Observation of Oxygen Vacancy Dynamics and Ordering in the Epitaxial LaCoO3 System. *Acs Nano* **2017,** *11* (7), 6942-6949.

26. Li, Q.; Nelson, C. T.; Hsu, S. L.; Damodaran, A. R.; Li, L. L.; Yadav, A. K.; McCarter, M.; Martin, L. W.; Ramesh, R.; Kalinin, S. V., Quantification of flexoelectricity in PbTiO3/SrTiO3 superlattice polar vortices using machine learning and phase-field modeling. *Nat. Commun.* **2017,** *8*.

27. Belianinov, A.; He, Q.; Kravchenko, M.; Jesse, S.; Borisevich, A.; Kalinin, S. V., Identification of phases, symmetries and defects through local crystallography. *Nat. Commun.* **2015,** *6*.

28. Yang, S.-Z.; Sun, W.; Zhang, Y.-Y.; Gong, Y.; Oxley, M. P.; Lupini, A. R.; Ajayan, P. M.; Chisholm, M. F.; Pantelides, S. T.; Zhou, W., Direct Cation Exchange in Monolayer $MoS_2$ via Recombination-Enhanced Migration. *Physical Review Letters* **2019,** *122* (10), 106101.

29. Yang, S.; Tian, X.; Wang, L.; Wei, J.; Qi, K.; Li, X.; Xu, Z.; Wang, W.; Zhao, J.; Bai, X.; Wang, E., In-situ optical transmission electron microscope study of exciton phonon replicas in ZnO nanowires by cathodoluminescence. *Applied Physics Letters* **2014,** *105* (7), 071901.

30. Choi, J.; Zhang, J.; Liou, S.-H.; Dowben, P. A.; Plummer, E. W., Surfaces of the perovskite manganites La 1− x Ca x MnO 3. *Phys. Rev. B.* **1999,** *59* (20), 13453.

31. Duerloo, K.-A. N.; Li, Y.; Reed, E. J., Structural phase transitions in two-dimensional Mo- and W-dichalcogenide monolayers. *Nature Communications* **2014,** *5*, 4214.

32. Eliseev, E. A.; Morozovska, A. N.; Glinchuk, M. D.; Zaulychny, B. Y.; Skorokhod, V. V.; Blinc, R., Surface-induced piezomagnetic, piezoelectric, and linear magnetoelectric effects in nanosystems. *Phys. Rev. B* **2010,** *82* (8), 085408.

33. Enyashin, A. N.; Yadgarov, L.; Houben, L.; Popov, I.; Weidenbach, M.; Tenne, R.; Bar-Sadan, M.; Seifert, G., New Route for Stabilization of 1T-WS2 and MoS2 Phases. *The Journal of Physical Chemistry C* **2011,** *115* (50), 24586-24591.

34. Eda, G.; Fujita, T.; Yamaguchi, H.; Voiry, D.; Chen, M.; Chhowalla, M., Coherent Atomic and Electronic Heterostructures of Single-Layer MoS2. *ACS Nano* **2012,** *6* (8), 7311-7317.

35. Lamfers, H. J.; Meetsma, A.; Wiegers, G. A.; de Boer, J. L., The crystal structure of some rhenium and technetium dichalcogenides. *Journal of Alloys and Compounds* **1996,** *241* (1), 34-39.

36. Estradé, S.; Arbiol, J.; Peiró, F.; Infante, I.; Sánchez, F.; Fontcuberta, J.; De la Peña, F.; Walls, M.; Colliex, C., Cationic and charge segregation in La 2/3 Ca 1/3 MnO 3 thin films grown on (001) and (110) SrTiO 3. *Appl. Phys. Lett.* **2008,** *93* (11), 112505.

37. Bühlmann, S.; Colla, E.; Muralt, P., Polarization reversal due to charge injection in ferroelectric films. *Phys. Rev. B.* **2005,** *72* (21), 214120.

38. Lin, Y.-C.; Komsa, H.-P.; Yeh, C.-H.; Björkman, T.; Liang, Z.-Y.; Ho, C.-H.; Huang, Y.-S.; Chiu, P.-W.; Krasheninnikov, A. V.; Suenaga, K., Single-Layer ReS2: Two-Dimensional Semiconductor with Tunable In-Plane Anisotropy. *ACS Nano* **2015,** *9* (11), 11249-11257.

39. Johari, P.; Shenoy, V. B., Tuning the Electronic Properties of Semiconducting Transition Metal Dichalcogenides by Applying Mechanical Strains. *ACS Nano* **2012,** *6* (6), 5449-5456.

40. Vlcek L.; Yang S.; Gong Y.; Ajayan P.; Zhou W.; Chisholm M. F.; Ziatdinov M.; Vasudevan R.K; S.V., K., Order and randomness in dopant distributions: exploring the thermodynamics of solid solutions from atomically resolved imaging. *arXiv preprint* **2019**, arXiv:1907.05531.
41. Lin, W. Z.; Li, Q.; Belianinov, A.; Sales, B. C.; Sefat, A.; Gai, Z.; Baddorf, A. P.; Pan, M. H.; Jesse, S.; Kalinin, S. V., Local crystallography analysis for atomically resolved scanning tunneling microscopy images. *Nanotechnology* **2013,** *24* (41).
42. Somnath, S.; Smith, C. R.; Laanait, N.; Vasudevan, R. K.; Ievlev, A.; Belianinov, A.; Lupini, A. R.; Shankar, M.; Kalinin, S. V.; Jesse, S., USID and Pycroscopy--Open frameworks for storing and analyzing spectroscopic and imaging data. *arXiv preprint arXiv:1903.09515* **2019**.
43. Somnath, S.; Smith, C. R.; Kalinin, S. V.; Chi, M.; Borisevich, A.; Cross, N.; Duscher, G.; Jesse, S., Feature extraction via similarity search: application to atom finding and denoising in electron and scanning probe microscopy imaging. *Advanced structural and chemical imaging* **2018,** *4* (1), 3.
44. AICrystallographer. https://github.com/pycroscopy/AICrystallographer.
45. Lin, W. Z.; Li, Q.; Belianinov, A.; Sales, B. C.; Sefat, A.; Gai, Z.; Baddorf, A. P.; Pan, M. H.; Jesse, S.; Kalinin, S. V., Local crystallography analysis for atomically resolved scanning tunneling microscopy images. *Nanotechnology* **2013,** *24*.
46. Rasmussen, C. E. In *The infinite Gaussian mixture model*, Advances in neural information processing systems, 2000; pp 554-560.
47. Rasmussen, C. E.; Williams, C. K. I., *Gaussian Processes for Machine Learning (Adaptive Computation and Machine Learning)*. The MIT Press: 2005.
48. Martin, O., *Bayesian Analysis with Python: Introduction to statistical modeling and probabilistic programming using PyMC3 and ArviZ, 2nd Edition*. Packt Publishing: 2018.
49. Lambert, B., *A Student's Guide to Bayesian Statistics*. SAGE Publications Ltd; 1 edition: 2018.
50. Kruschke, J., *Doing Bayesian Data Analysis: A Tutorial with R, JAGS, and Stan*. Academic Press; 2 edition: 2014.
51. Dyck, O.; Kim, S.; Kalinin, S. V.; Jesse, S., Placing single atoms in graphene with a scanning transmission electron microscope. *Appl. Phys. Lett.* **2017,** *111* (11).
52. Dyck, O.; Kim, S.; Jimenez-Izal, E.; Alexandrova, A. N.; Kalinin, S. V.; Jesse, S., Building Structures Atom by Atom via Electron Beam Manipulation. *Small* **2018,** *14* (38).
53. Toma Susi, J. C. M., Jani Kotakoski, Manipulating low-dimensional materials down to the level of single atoms with electron irradiation. *Ultramicroscopy* **2017,** *180*, 163–172.

**Supplementary Materials**

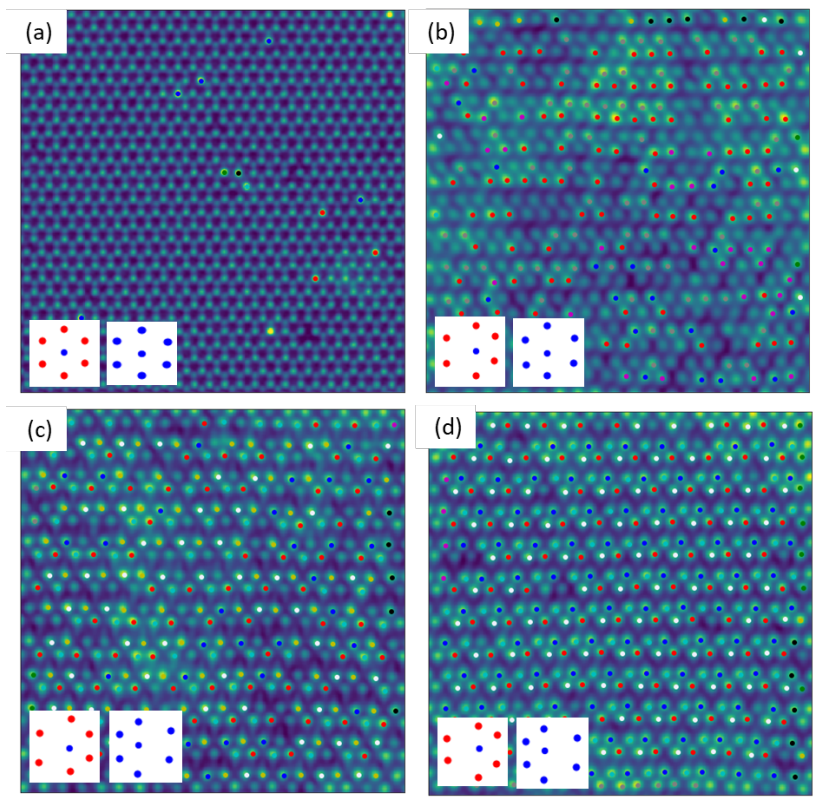

**Figure S1.** Clustering analysis of the structure of the of $Mo_{1-x}Re_xS_2$ with $x = 0.05$, $0.55$, $0.78$, and $0.95$ (a-d) using *k*-means clustering algorithm on the local neighborhood. Only two types of clusters with Re as the central atom are shown; further clusters are visualized in the given Jupyter notebook.

A complementary representation for the internal structure can be obtained via the *k*-means clustering analysis of the atomic neighborhood. While the PCA analysis provides the information on the symmetry distortions from the high-symmetry state, the *k*-mean clustering groups the atoms based on the similarity of local structural motifs. Shown in Fig. S1 are the cluster structures (insets) and their location for the four different compositions. For simplicity only two clusters are shown; there are more (see Jupyter notebook). The features on which the separation is done are the distances to the nearest neighbors. Given that there is an order to the neighbors, rotational variants will show up as distinct clusters in this analysis.

Here again one of the preponderant cluster groups is the one corresponding to the displacement of a single neighbor pair relative to the central atom, and its orientational variants. Additional clusters are formed as a result of the indexing errors, when the nearest atom is missing and therefore *k*-means vector centers around one of the next-nearest atoms. Note that for ordered phases, the clusters of the similar atoms start to adopt the periodic structure, as expected for the long-range ordered material. At the same time, for the intermediate doping, the clusters are (visually) random distributed. However, this information provides the insight into the distributions of local structural distortions that define atomic mechanisms of phase transitions.

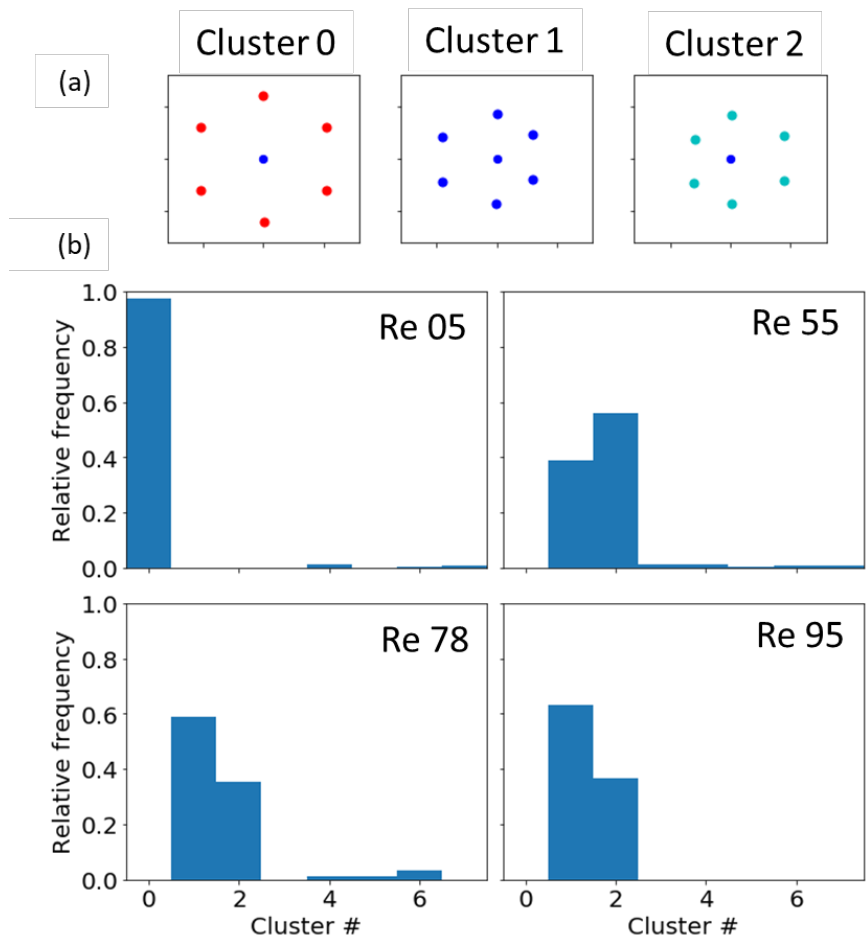

**Figure S2.** K-means analysis across the phase transition for the full composition series. Note that the phase evolution can be represented via relative changes in 3 preponderant structural elements, where cluster 0 is maximal for the H phase and clusters 1 and 2 appear for the distorted structures. Note that intermediate compositions also contain other structural elements, representing relatively higher disorder compared to the other cluster means.

For insight into the evolution of the local structure across the phase transition, we perform a crystallographic analysis with clustering across the full composition series. Here, the full dictionary of atomic configurations for all composition is statistically analyzed. The analysis is shown in Fig. S2, showing the dominant structural distortions. Unsurprisingly, the first composition with the least amount of Re contains the dominant hexagonal symmetry characteristic of MoS$_2$. Very dilute traces of other clusters can be seen. For the x=0.55 composition and onwards, the clusters shift towards clusters 1 and 2, which are characteristic of the ReS$_2$ dominant (triclinic) symmetry.

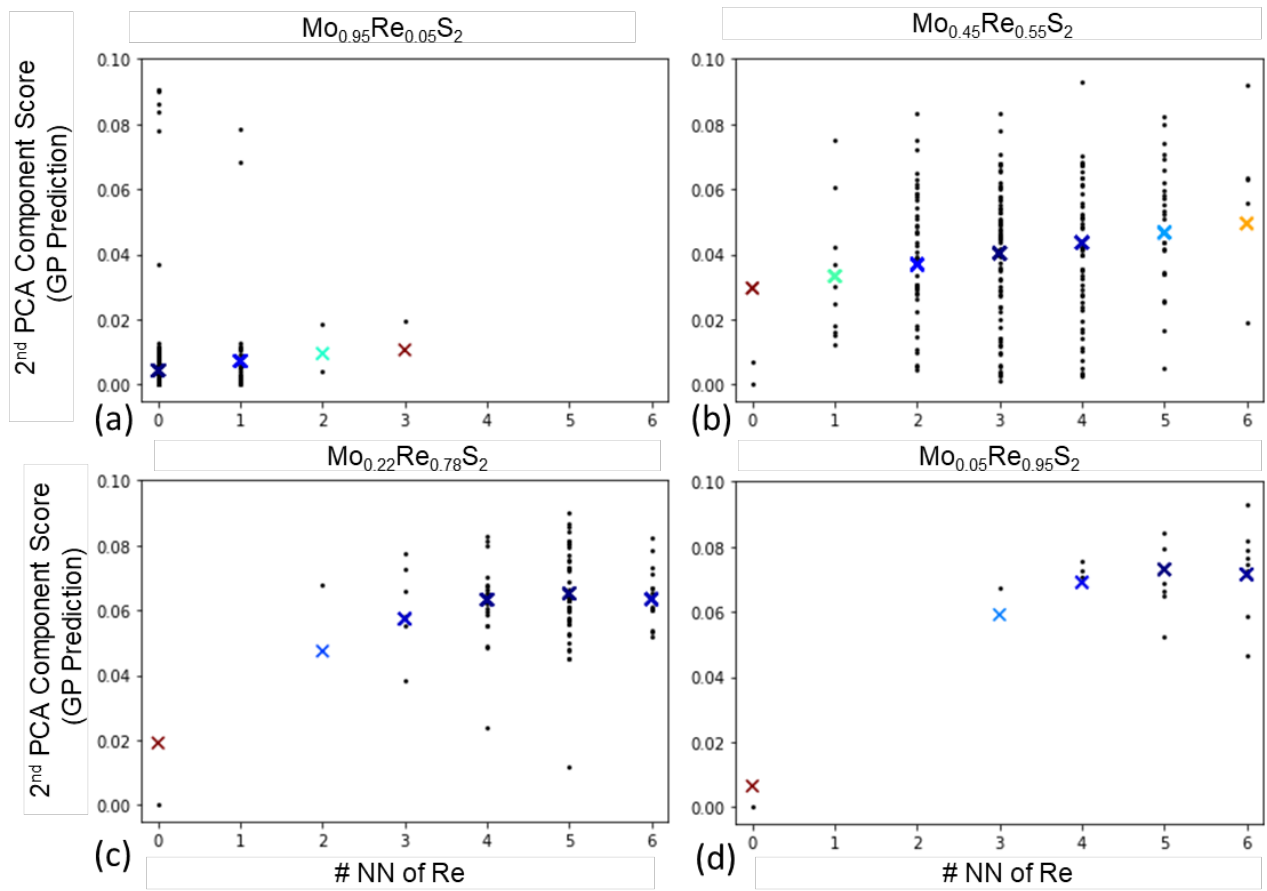

**Figure S3.** Gaussian Process averaged 2$^{nd}$ PCA component scores via local concentration for 4 explored compositions.